\pdfoutput=1
\documentclass[
 reprint,
 amsmath,amssymb,
 aps,
 nofootinbib,
 superscriptaddress,
 longbibliography,
 preprintnumbers
]{revtex4-2}

\usepackage{amsthm}
\usepackage{graphicx}
\usepackage{xcolor}
\usepackage{braket}
\usepackage{amsmath}
\usepackage{amssymb}
\usepackage{enumerate}
\usepackage{mathtools}
\usepackage{etoolbox}
\usepackage{algorithmic}
\usepackage{algorithm}
\usepackage{tikz}
\usepackage{bbm}
\usepackage{here}
\usepackage{subfigure}
\usepackage[
colorlinks=true,
urlcolor=blue,
citecolor=blue,
linkcolor=blue,
hyperfootnotes=false,
breaklinks=true
]{hyperref}
\usepackage{comment}
\usepackage{makecell}
\usepackage{multirow}
\usepackage{ulem}

\usetikzlibrary{quantikz2}

\makeatletter
\renewcommand{\p@subsection}{}
\renewcommand{\p@subsubsection}{}
\makeatother

\newcommand{\aap}{Astronomy \& Astrophysics}

\newcommand{\physrep}{Physics Reports}

\allowdisplaybreaks

\begin{document}
\preprint{KEK-TH-2677, {\it prepared for submission to Computers and Fluids}}
\title{Quantum simulation of Burgers flow: Nonlinear transformation and direct evaluation of statistical quantities}

\author{Fumio Uchida}
\email{fumiouchida@ibs.re.kr}
\affiliation{Theory Center, Institute of Particle and Nuclear Studies (IPNS), High Energy Accelerator Research Organization (KEK), Tsukuba, Ibaraki 305-0801, Japan}
\affiliation{Kavli IPMU, WPI, University of Tokyo, 5-1-5 Kashiwanoha, Kashiwa, Chiba 277-8583, Japan}
\affiliation{CTPU-CGA, Institute for Basic Science (IBS), Daejeon, 34126, Korea}

\author{Koichi Miyamoto}
\email{miyamoto.kouichi.qiqb@osaka-u.ac.jp}
\affiliation{Center for Quantum Information and Quantum Biology, University of Osaka, Toyonaka, Osaka 560-0043, Japan}

\author{Soichiro Yamazaki}
\email{soichiro.yamazaki@phys.s.u-tokyo.ac.jp}
\affiliation{Department of Physics, University of Tokyo, 7-3-1 Hongo, Bunkyo, Tokyo 113-0033, Japan}

\author{Kotaro Fujisawa}
\email{kotaro.fujisawa@gmail.com}
\affiliation{Department of Liberal Arts, Tokyo University of Technology, 5-23-22 Nishikamata, Ota, Tokyo, 144-8535, Japan}
\affiliation{Research Center for the Early Universe (RESCEU), 7-3-1 Hongo, Bunkyo, Tokyo, 113-0033, Japan}

\author{Naoki Yoshida}
\email{naoki.yoshida@ipmu.jp}
\affiliation{Kavli IPMU, WPI, University of Tokyo, 5-1-5 Kashiwanoha, Kashiwa, Chiba 277-8583, Japan}
\affiliation{Department of Physics, University of Tokyo, 7-3-1 Hongo, Bunkyo, Tokyo 113-0033, Japan}
\affiliation{RIKEN Center for Advanced Intelligence Project, 1-4-1 Nihonbashi, Chuo, Tokyo 103-0027, Japan}

\date{\today}

\begin{abstract}
\noindent
Fault-tolerant quantum computing is a promising technology to solve linear partial differential equations that are classically demanding to integrate. It is still challenging to solve nonlinear equations in fluid dynamics, such as the Burgers equation, using quantum computers. We propose a novel quantum algorithm to solve the Burgers equation. With the Cole--Hopf transformation that maps the fluid velocity field $u$ to a new field $\psi$, we apply a sequence of quantum gates to solve the resulting linear equation and obtain the quantum state $\ket{\psi}$ that encodes the solution $\psi$.
We also propose an efficient way to extract stochastic properties of $u$, namely the multi-point functions of $u$, from the quantum state of $\ket{\psi}$.
Our algorithm offers an exponential advantage over the classical finite difference method in terms of the number of spatial grids when a perturbativity condition in the information-extracting step is met. 
\end{abstract}

\maketitle

\section{Introduction}
\noindent
With the recent rapid development of hardware platforms, quantum computing attracts great interest and hope in physical science \cite{Jordan:2012xnu,Jordan:2011ci,Jordan:2014tma,Jordan:2017lea,Preskill:2018fag,Bauer:2022hpo,DiMeglio:2023nsa,Delgado:2022tpc}, including cosmology and astrophysics \cite{Kaufman:2019vzn,Liu:2020wtr,Joseph:2021naq,Czelusta:2021jro,Mocz:2021ehj,Yamazaki:2023emt,Miyamoto:2023iwk,Higuchi:2024nyt}.
Although we are still in the era of noisy intermediate-scale quantum technologies (NISQ) \cite{Preskill:2018jim}, there have been a variety of studies aimed at fault-tolerant quantum computing (FTQC) \cite{Shor:1996qc,Campbell:2017esv,Webster:2020xsa,GoogleQuantumAI:2022fyn}, which enables us to tackle classically intractable problems.
Fluid dynamics continues to provide such challenging problems.
For instance, Refs.~\cite{2020JCoPh.40909347T,Yamazaki:2023emt,Miyamoto:2023iwk,Higuchi:2024nyt} proposed quantum algorithms to integrate the collisionless Boltzmann equation (CBE) \cite{1982A&A...114..211H}, which describes the dynamics of a dilute system such as rarefied gas, and cosmic neutrinos and dark matter in the universe.
Reference~\cite{2025PhRvA.112a0401O} proposed a quantum simulation of passive scalar dynamics in a given fluid velocity field.

There remains an ``apparent'' limitation that quantum computing is applicable to only linear problems in a direct manner because of the linearity of quantum operations. 
Although quantum algorithms to solve a wide class of linear problems have been proposed \cite{Harrow:2008dvn,Clader:2013gxx,Berry:2014jeh,AndrewM:2015yvx,Costa:2019wdc,Sato:2024ihp}, solving nonlinear problems has been a challenging task for FTQC algorithms.

Nevertheless, applying quantum computing to nonlinear problems such as the Navier--Stokes equation (NSE) is important and timely.
The NSE is a nonlinear partial differential equation that governs fluid dynamics \cite{1987flme.book.....L}, and there are a broad range of potential applications in astrophysics and cosmology, {\it e.g.}, the primordial gravitational wave \cite{RoperPol:2019wvy,Brandenburg:2023imm,Sharma:2022ysf} and the primordial magnetic field \cite{Brandenburg:1996fc,Banerjee:2004df,Kahniashvili:2012uj,Brandenburg:2017neh,Uchida:2024ude} coupled with a photon-baryon plasma in the early universe, as listed in Ref.~\cite{2016arXiv160403835O}.
Although no general framework is known to efficiently solve the NSE, a variety of attempts have been proposed \cite{2008arXiv0812.4423L,2020npjQI...6...61G,Meng:2023zek,Oz:2021dka,2019arXiv190409033R,2024arXiv240618749S}.
Reference~\cite{2025NatRP...7..220T} provides a comprehensive overview of this topic.

Several approaches have been proposed to deal with nonlinear problems with FTQC algorithms. 
Many of them are based on linearization of the nonlinear equations by introducing a large number of auxiliary variables, namely the Carleman embedding and its generalizations \cite{2021PNAS..11826805L,2021PhPl...28f2305E,2023Quant...7..913K}.\footnote{
    Another class of linearization method includes Koopman--von-Neumann approach \cite{2020PhRvR...2d3102J} and Fokker--Planck approach \cite{2024arXiv240113500T,2025RSPSA.48150016T}, which solve the evolution of distribution function in the phase space. Unfortunately, the infinite dimension of the phase space can be an obstacle when one wishes to simulate the dynamics of field variables.}
Practically, we need to truncate auxiliary variables at a finite number and solve the truncated problem with quantum computing.
The accuracy of this procedure is guaranteed only for problems with moderate non-linearity.

An intriguing problem is the one-dimensional Burgers equation \cite{1915MWRv...43..163B,burgers1948mathematical}, which appears, for example, as a model of traffic flow \cite{nagatani2002physics}.
It is a nonlinear equation that describes the time evolution of fluid velocity $u$, and can be viewed as the simplest model of nonlinear fluid dynamics and hence as an important first step to solve the full NSE \cite{2012PhFl...24e5113O,2024arXiv240806529O}.
A crucial point is that the one-dimensional Burgers equation is transformed to a linear heat equation by the Cole--Hopf transformation \cite{hopf1950partial,cole1951quasi}, which maps $u$ to a new field $\psi$.
However, this does not mean that Burgers turbulence is completely understood. 
Turbulent velocities are described by the Burgers equation, and its stochastic properties have been extensively investigated \cite{1966PhFl....9.2114J,1993PhFlA...5..445G,1998CMaPh.191...71R,1999PhFl...11.2143G,2022PhRvF...7g4605A,2007PhR...447....1B}.\footnote{
    That being said, understanding turbulence based on the Navier--Stokes equation is a different problem. Note that, although the emergence of shocks in the solution is a distinctive feature of the Burgers equation in the inviscid limit \cite{2000nlin.....12033F}, $\nu\to0$ in Eq.~\eqref{eq:Burgers}, we keep $\nu$ to be finite so that the Cole--Hopf transformation works. Our aim here is to simulate the Burgers turbulence with finite $\nu$ and hence without discontinuous shocks.}

In the present study, we propose a quantum algorithm to measure the statistical properties of the one-dimensional Burgers flow.
We apply the quantum differential equation solver \cite{An:2023qss} to the equation obtained by the Cole--Hopf transformation on the Burgers equation to get the quantum state $\ket{\psi}$ that encodes the solution $\psi$ in the amplitudes.
Furthermore, from $\ket{\psi}$, we extract the statistical properties of the solution, specifically, the multi-point functions of $u$, via estimating expectation values of some operators.
This is a well-defined and well-suited task for quantum computing.
If one would like to understand the full real-space configuration, measurement with a given accuracy would cost an enormous time complexity.
In contrast, we aim at deriving only a limited number of values as classical data from a quantum state, taking full advantage of a quantum speedup.
In this information extraction step, due to the nonlinearity of the Cole--Hopf transformation, we need to apply a certain approximation that a perturbative nature holds with respect to $\psi$, which is non-trivial for fluid with a large Reynolds number.
Nevertheless, when the perturbative condition holds, our algorithm offers an exponential advantage over the classical finite difference method in terms of the number of spatial grids.
To validate our approach, we perform simple test calculations and classically show that the solution by our method is sufficiently accurate compared to that without the approximation.

The paper is organized as follows.
In Sec.~\ref{sec:NG}, we introduce the Burgers equation and multi-point functions of the velocity field.
In Sec.~\ref{sec:QA}, we present our algorithm to solve the Burgers equation and extract the multi-point functions from the solution-encoding quantum state.
In Sec.~\ref{sec:Discussion}, we discuss the complexity analysis for our algorithm and possible generalizations, along with presenting the results of numerical demonstrations.
Sec.~\ref{sec:Conclusion} concludes this paper.

\section{Burgers flow\label{sec:NG}}
\subsection{Burgers equation and Cole--Hopf transformation}
\noindent
We consider a velocity field $u(t,x)$ in one-dimensional system, which is governed by the Burgers equation
\begin{align}
    \partial_t u +u \partial_x u
        =\nu \partial^2_x u,
    \label{eq:Burgers}
\end{align}
where $\nu>0$ is a viscosity coefficient that dissipates kinetic energy into heat.
The advection term introduces a nonlinearity, of which significance may be quantified as the Reynolds number,
\begin{align}
    {\rm Re}
        \coloneq \dfrac{ul}{\nu},
    \label{eq:Re_def}
\end{align}
where $l$ is the typical length scale of the system, and ${\rm Re}\gg1$ implies a highly nonlinear dynamics.

However, irrespectively of how large the Reynolds number is, Hopf \cite{hopf1950partial} and Cole \cite{cole1951quasi} found that, by introducing a new field $\psi$, {\it s.t.}
\begin{align}
    \psi
        =\exp\left(-\frac{1}{2\nu}\int^xdy\,u(y)\right),
    \label{eq:CHtrsf}
\end{align}
or equivalently,
\begin{align}
    u
        =-2\nu\dfrac{\partial_x\psi}{\psi}
        \label{eq:psi_to_u},
\end{align}
Eq.~\eqref{eq:Burgers} is reduced to a linear heat equation
\begin{align}
    \partial_t\psi
        =\nu\partial_x^2\psi.
        \label{eq:heateqn}
\end{align}

The heat equation can be formally integrated, although its numerical evaluation is not necessarily easy.
Our aim is to bypass the {\it nonlinearity} of the original problem with the Cole--Hopf transformation and take advantage of the {\it linear} nature of quantum computing.
Note that solving the heat equation with a quantum speedup is an area of study \cite{2022CMaPh.395..601L}, but we do not extensively investigate this point in this work.

\subsection{Statistical quantities of Burgers turbulence}
\noindent
We suppose that we are interested in statistical quantities, namely the multi-point functions, of the velocity field.
We define the multi-point functions $P^{(n)}$ and the higher-order moments $I^{(n)}$ for $n\geq2$ of the velocity field
\begin{align}
    P^{(n)}\left(\{x_i-x_0\}_{i=1,\cdots n-1},t\right)
        \coloneq\overline{\prod_{i=0}^{n-1} u(x_i,t)},\\
    I^{(n)}(t)
        \coloneq P^{(n)}\left(\{0\}_{i=0,\cdots n-1},t\right),
\end{align}
where the bar takes a volume average, where $\overline{X(\{x_i\}_i)}\coloneq (1/L)\int_0^Ldx'\,X(\{x'+x_i\}_i)$ for a general function $X$ with the periodic boundary condition, of the inside.
We may further average over ensembles, when we have a bunch of random initial conditions, to explore the universal properties of Burgers turbulence with those initial conditions.

In the following sections, we will propose an algorithm to measure these quantities up to a common normalization factor among all the quantities.

\section{Quantum algorithm\label{sec:QA}}
\noindent
In this section, we will explain that the Burgers equation can be solved efficiently, by using a quantum computer.

The algorithm to compute the multi-point functions for one realization of the velocity field proceeds in the following three steps:
\begin{itemize}
    \item Classical to quantum:\\ loading the initial condition into the quantum state (Sec.~\ref{sec:C2Q})
    \item Quantum operation:\\integration of the heat equation (Sec.~\ref{sec:Q})
    \item Quantum to classical:\\extracting the wanted information (Sec.~\ref{sec:Q2C})
\end{itemize}

On top of this, in Sec.~\ref{sec:En}, we explain how to efficiently take an ensemble average of the multi-point functions.

\subsection{Implementing the initial condition\label{sec:C2Q}}
\noindent
We discretize the spatial coordinate $0\leq x\leq L$ into $N_x=2^{n_x}$ grids and impose, for now, the periodic boundary condition.
Generalization to non-periodic conditions will be discussed in Sec.~\ref{sec:Discussion}.
We assume that we have an oracle $O_{\psi'_0}$ to generate a quantum state encoding the initial value of $\partial_x \psi$:
\begin{align}
    O_{\psi'_0}\left\vert\boldsymbol{0}\right\rangle
        =\left\vert\partial_x\boldsymbol{\psi}(0)\right\rangle.
\end{align}
Here, we define
\begin{align} 
    \left\vert\partial_x\boldsymbol{\psi}(t)\right\rangle
        &\coloneq\sum_{j=0}^{N_x-1}\dfrac{\partial_x\psi_j(t)}{\mathcal N(t)}\,\left\vert j\right\rangle
    \label{eq:iniCondState}
\end{align}
where $\partial_x\psi_j(t)\coloneq\partial_x\psi(j\Delta x,t),$ and $\Delta x\coloneq L/N_x$.
A normalization factor $\mathcal N$ is introduced so that $\langle\partial_x\boldsymbol{\psi}(t)\vert\partial_x\boldsymbol{\psi}(t)\rangle=1$.

The methods to generate a quantum state encoding a known function like Eq. \eqref{eq:iniCondState} have been investigated in the literature.
For example, if $\partial_x\psi(x,0)$ is analytically calculated and satisfies some conditions, the Grover-Rudolph method \cite{grover2002creating} and other methods \cite{Sanders2019,rattew2022preparing,mcardle2022quantum,MarinSanchez2023,Moosa_2024} can be used.
If $\partial_x\psi_j(0)$ is given pointwise and the values are stored in the quantum random access memory \cite{2008PhRvL.100p0501G}, we can generate $\ket{\partial_x\boldsymbol{\psi}(0)}$ by the method in Ref.~\cite{Kerenidis2020}.

\subsection{Integration of the heat equation\label{sec:Q}}
\noindent
Let us discretize the heat equation \eqref{eq:heateqn}.
In terms of $\partial_x\psi$, Eq.~\eqref{eq:heateqn} is
\begin{align}
    \partial_t\partial_x\psi
        =\nu\partial_x^2\partial_x\psi.
        \label{eq:d_heateqn}
\end{align}
By normalizing the temporal coordinate properly, namely $\tau\coloneq\nu t/\Delta x^2$, and discretizing the spatial coordinate with the central differencing scheme\footnote{
    We demonstrate the applicability of this scheme by explicitly reproducing characteristic solutions of the Burgers equation in Appendix \ref{sec:demo}.}, it becomes
\begin{align}
    \dfrac{d\partial_x\psi_j}{d\tau}
        =\partial_x\psi_{j-1}-2\partial_x\psi_j+\partial_x\psi_{j+1},\quad
        j\in {\mathbb Z}_{N_x}.
        \label{eq:discretized}
\end{align}
Equation \eqref{eq:discretized} can also be written as
\begin{align}
    \dfrac{d\partial_x\boldsymbol{\psi}}{d\tau}
        =A\partial_x\boldsymbol{\psi},\quad
    A
        \coloneq\left(\begin{matrix}
            -2&1&0&\cdots&1\\
            1&-2&1&\cdots&0\\
            0&1&-2&\cdots&0\\
            &&\cdots&&\\
            1&0&\cdots&1&-2
        \end{matrix}\right),
    \label{eq:defA}
\end{align}
which is integrated formally as
\begin{align}
    \partial_x\boldsymbol{\psi}(\tau)
        =e^{A\tau}\partial_x\boldsymbol{\psi}(0).
    \label{eq:formalintegration}
\end{align}
Note that small values for $\nu$ can approximate the shock solution in the inviscid Burgers equation,  requiring us to take small $\Delta x$ and hence large $N_x\propto \nu^{-1/2}$, accordingly.

Obtaining $\left\vert\partial_x\boldsymbol{\psi}(\tau)\right\rangle$ as a quantum state is possible by the method presented in Ref.~\cite{An:2023qss}, whose outline is as follows.
The point is block-encoding of $e^{A\tau}$, which is achieved by the technique called the linear combination of Hamiltonian simulation.
Since $A$ is negative semi-definite with eigenvalues $2(\cos(2\pi k/N_x)-1), k=0,\cdots N_x-1$ \cite{gray2006toeplitz}, we have a decomposition formula
\begin{align}
    e^{A\tau}=\int_{-\infty}^{\infty}\dfrac{f(k)}{-2\pi i(k+i)}e^{ikA\tau}dk,
    \label{eq:her2uni}
\end{align}
where $f(z):=e^{(1+iz)^\beta-2^\beta}$ with an appropriately chosen constant $\beta$ \cite{An:2023qss}.
The integral in Eq.~\eqref{eq:her2uni} can be approximated by a piecewise Gaussian quadrature, consequently as
\begin{align}
    \text{Eq.~\eqref{eq:her2uni}}\simeq\sum_m c_m e^{+ik_mA\tau},
    \label{eq:disc}
\end{align}
where $c_m$s are $\mathcal O(1)$ weights.
Since $e^{ik_mA\tau}$ is unitary, Eq. \eqref{eq:disc} can be block-encoded via the method called linear combination of unitaries \cite{childs2012hamiltonian,PhysRevLett.114.090502}.
According to Corollary 16 of Ref.~\cite{An:2023qss}, since $\|A\|\le \sqrt{\|A\|_1 \|A\|_\infty}=4$~\footnote{For $M=(m_{ij})\in\mathbb{C}^{m \times n}$, $\|M\|$ denotes its spectral norm, $\|M\|_1=\max_{j}\sum_{i=1}^m |m_{ij}|$, and $\|M\|_\infty=\max_{i}\sum_{j=1}^n |m_{ij}|$.}, an oracle $O_{\psi'_\tau}$ that act on the initial state $\left\vert\partial_x\boldsymbol{\psi}(0)\right\rangle$ to generate $\vert\partial_x\boldsymbol{\psi}(\tau)\rangle$ with an error tolerance $\epsilon_1>0$ costs
\begin{align}
    \tilde{\mathcal O}\left(\dfrac{\vert\vert\partial_x\boldsymbol\psi_0\vert\vert}{\vert\vert\partial_x\boldsymbol\psi(\tau)\vert\vert}\tau\times \mathrm{polylog}\left(\dfrac{1}{\epsilon_1}\right)\right)\;\text{\it queries to}\; U_A,\label{eq:TEquery1}\\
    \text{\it and }\;{\mathcal O}\left(\dfrac{\vert\vert\partial_x\boldsymbol\psi_0\vert\vert}{\vert\vert\partial_x\boldsymbol\psi(\tau)\vert\vert}\right)\;\text{\it queries to}\; O_{\psi'_0},\label{eq:TEquery2}
\end{align}
where $U_A$ is a block-encoding of $A$. 
If the block-encoding of $A$ has an error $\epsilon_2$, the factor $e^{A\tau}$ in Eq.~\eqref{eq:formalintegration} has a relative error $\epsilon_2\tau$, resulting in a total error $\epsilon_1+\epsilon_2\tau$ for the state $\vert\partial_x\boldsymbol{\psi}(\tau)\rangle$.
According to Ref.~\cite{gilyen2019quantum} (Lemma 48 in the full-version), with $A$'s sparsity and max norm being 3 and 2, respectively, we can implement a $(6,n_x+3,\epsilon_2)$ block-encoding\footnote{For the definition of $(\alpha,a,\epsilon)$ block-encoding, where $\alpha,\epsilon>0$ and $a\in\mathbb{N}$, see Ref.~\cite{gilyen2019quantum}.} of $A$, using
\begin{align}
    {\mathcal O}\left(n_x+\log^{\frac{5}{2}}\left(\dfrac{18}{\epsilon_2}\right)\right)\;{\text{\it fundamental gates}}\label{eq:BEquery1}   
\end{align}
and $\mathcal{O}(\log^{5/2}(18/\epsilon_2))$ ancilla qubits, along with
\begin{align}
    \mathcal{O}(1) ~ {\text{\it calls to the sparse-access oracles for }} A.
    \label{eq:BEquery2}
\end{align}

We will discuss query complexity of our algorithm in terms of the numbers of uses of the sparse-access oracles, which access elements of a sparse matrix, and fundamental gates, {\it i.e.}, a universal set of one- and two-qubit gates, not included therein.

At this point, we have an oracle $O_{\psi'_\tau}O_{\psi'_0}$ to generate the ``solution" state $\left\vert\partial_x\boldsymbol{\psi}(\tau)\right\rangle$, which has values of $\partial_x\psi_j(\tau)$ in the amplitude of $\vert j\rangle$.
Nonlinearity in terms of $u$ and high spatial resolution is achieved up to limitations originating from the finite discretization scheme \cite{1991JCoPh..97...73M}.

However, extracting arbitrary information from a quantum state is a non-trivial task, and we address this issue in the next section.

\subsection{Retrieving the desired information \label{sec:Q2C}}
\noindent
In this step, we introduce approximations to extract statistical information on the velocity field $u$.
We begin by discussing the two- and three-point functions $P^{(2)}$ and $P^{(3)}$, and then generalize them to higher-order multipoint functions.

\subsubsection{Two-point function}
\noindent
We apply a perturbative description to linearize the relation between quantum states that encode the solution $\psi(t)$ in their amplitudes and $u$.
We approximate the velocity field $u$ as
\begin{align}
    u_j
        \simeq
        -\dfrac{\nu \partial_x \psi_j}{\tilde\psi_j},
    \label{eq:dpsi_to_u}
\end{align}
where we discretize Eq.~\eqref{eq:psi_to_u} and replace $\psi$ in the denominator with an approximation $\tilde\psi$. Note that we have not introduced any approximation in the procedure of time-integration of the Burgers equation.
Instead, the only approximation is made at the information extraction stage.
This is a novel approach to cope with the nonlinearity, differing from existing ones that focus on approximate linearizations of fluid time-evolution equations, e.g., via Carleman linearization.
Explicit examples of $\tilde\psi$ will be discussed later in Sec.~\ref{sec:Discussion}.
Hereafter, we suppose that we know $\tilde\psi$ independently of the solution state $\vert\partial_x\boldsymbol\psi\rangle$, which we call as the zero-th order approximation, and consider the statistics on the right-hand side of Eq.~\eqref{eq:dpsi_to_u}.
With this approximation, we have
\begin{align}
    \left\vert\boldsymbol u\right\rangle
        &\coloneq\sum_j\dfrac{u_j}{\vert\vert\boldsymbol u\vert\vert}\vert j\rangle\notag\\
        &\simeq-
        \Lambda
        \left\vert\partial_x\boldsymbol{\psi}(t)\right\rangle,
    \label{eq:approx_u_partialpsi}
\end{align}
where $\Lambda$ is a diagonal operator that gives a weight to each component of the vector $\partial_x\boldsymbol\psi$, such that
\begin{align}
    \Lambda\left\vert\partial_x\boldsymbol{\psi}\right\rangle
        =\sum_{j=0}^{N_x-1}\dfrac{\partial_x\psi_j}{\mathcal N' \tilde\psi_j}\,\left\vert j\right\rangle,
\end{align}
where $\mathcal N'$ is a $j$-independent normalization factor such that $\langle \partial_x\boldsymbol{\psi}\vert \Lambda^\dagger\Lambda\vert\partial_x\boldsymbol{\psi}\rangle=1$.
In pariticular, $\Lambda$ is an identity transformation and $\mathcal N'\tilde\psi=\mathcal N$ when we take a homogeneous $\tilde\psi=\bar\psi\coloneq\frac{1}{L}\int_0^L\psi(x,t) dx$.
Otherwise, for a given error tolerance $\epsilon_\Lambda>0$, we can construct $U_\Lambda$, an $(a,n_x+3, \epsilon_\Lambda)$ block encoding of $\Lambda$, using $\mathcal O (n_x+\log^{5/2}(a/\epsilon_\Lambda))$ fundamental gates and $\mathcal O (\log^{5/2}(a/\epsilon_\Lambda))$ ancilla qubits.
Here, we have defined $a\coloneq\mathcal N/(\mathcal N'\tilde\psi_{\rm min})\leq\tilde\psi_{\rm max}/\tilde\psi_{\rm min}$, which is bounded independently of $N_x$, where $\tilde\psi_{\rm max}\coloneq\max_j\tilde\psi_j$ and $\tilde\psi_{\rm min}\coloneq\min_j\tilde\psi_j$.

Then, the two-point function of the velocity field $P^{(2)}(r=\rho\Delta x)$, $\rho\in{\mathbb Z}_{N_x}$ can be approximated as
\begin{align}
    P^{(2)}\left(r\right)
        &=\overline{u(x)u(x+r)}\notag\\
        &=N_x^{-1}\sum_{k=0}^{N_x-1}u_{k}u_{k+\rho}\notag\\
        &=N_x^{-1}\vert\vert\boldsymbol u\vert\vert^2\sum_{j,k=0}^{N_x-1}C^{(2)}_{jk}(\rho)\dfrac{u_j}{\vert\vert\boldsymbol u\vert\vert}\dfrac{u_k}{\vert\vert\boldsymbol u\vert\vert}.
        \label{eq:2pt_matrixform}
\end{align}
Here, $C^{(2)}_{jk}(\rho)$ is given by
\begin{align}
    C^{(2)}_{jk}(\rho)
        &\coloneq\sum_{\sigma,\sigma'}\delta_{j-\sigma,k-\sigma'}c^{(2)}_{\sigma\sigma'}(\rho),
\end{align}
where
\begin{align}
    c^{(2)}_{\sigma\sigma'}(\rho)
        &\coloneq\left\{\begin{matrix}
        +\dfrac{1}{2}&\begin{matrix}
            (\sigma,\sigma')=(0,\rho),\\
            (\sigma,\sigma')=(\rho,0),
        \end{matrix}\\
        0&\text{otherwise}
        \end{matrix}\right.
\end{align}
for $\rho\neq0$, and
\begin{align}
    c^{(2)}_{\sigma\sigma'}(0)
        \coloneq\left\{\begin{matrix}
        1& \sigma=\sigma'=0, \\
        0&\text{otherwise}
        \end{matrix}\right.
\end{align}
for $\rho=0$.
From Eq.~\eqref{eq:2pt_matrixform}, we obtain an expression for the two-point function $P^{(2)}(r,\tau)$ normalized by the second moment $I^{(2)}(\tau)=P^{(2)}(0,\tau)$,
\begin{align}
    \dfrac{P^{(2)}\left(r,\tau\right)}{I^{(2)}(\tau)}
        =\dfrac{\left\langle{\boldsymbol{u}}(\tau)\vert C^{(2)}(\rho)\vert{\boldsymbol{u}}(\tau)\right\rangle}{\left\langle{\boldsymbol{u}}(\tau)\vert C^{(2)}(0)\vert{\boldsymbol{u}}(\tau)\right\rangle}.
    \label{eq:P2/I2}
\end{align}
Both the numerator and the denominator are $\mathcal O(1)$ quantities, as understood from the expression in Eq.~\eqref{eq:2pt_matrixform}, where $N_x^{-1}\vert\vert\boldsymbol u\vert\vert^2=\mathcal O(1)$.

The values of the numerator and the denominator in the right-hand side can be measured by applying the overlap estimation algorithm \cite{2007PhRvA..75a2328K}.
For that purpose, we decompose $C^{(2)}(\rho)$ as
\begin{align}\hspace{-2.5mm}
    C^{(2)}(\rho)
        =\dfrac{1}{2}\left(P_{N_x}^{\rho}+P_{N_x}^{-\rho}\right),
    \label{eq:C2_decomp}
\end{align}
where $P_{N_x}$ is the cyclic increment
\begin{align}
    P_{N_x}
        \coloneq\sum_{j=0}^{N_x-2} \ket{j+1}\bra{j} + \ket{0}\bra{N_x-1},    
\end{align}
which is impremented with ${\mathcal{O}}(n_x)$ fundamental gates and ancilla qubits \cite{li2014class}.
Therefore, for $\rho=\sum_{a=0}^{n_x} \rho_a 2^{a}$, because we can decompose $P_{N_x}^{\rho}$ as $P_{N_x}^{\rho}=\prod_{a=0}^{n_x} \delta_{\rho_a,1}P_{2^{n_x-a}}\otimes 1_{2^{a}}$, we can construct $P_{N_x}^{\rho}$ with
\begin{align}
{\mathcal{O}}(n_x^2)~{\text{\it fundamental gates}}
\label{eq:C2GateCost}
\end{align}
and ${\mathcal{O}}(n_x)$ ancilla qubits, which is reused in each $P_{2^{n_x-a}}$.

Since $P_{N_x}^{\rho}$ is unitary, we can estimate $\bra{\boldsymbol{u}(\tau)} P_{N_x}^{\rho}\ket{\boldsymbol{u}(\tau)}$ by the overlap estimation algorithm in \cite{2007PhRvA..75a2328K}, making
\begin{align}
    {\mathcal{O}}\left(\dfrac{1}{\epsilon_3}\right)\text{\it queries to}\; P_{N_x}^{\pm\rho},\;P_{N_x}^{\pm\rho\pm2},\text{\it and }\; O_{u_\tau},
\end{align}
where $O_{u_\tau}\coloneq-O_{\psi'_\tau}O_{\psi'_0}$, to achieve an error tolerance $\epsilon_3>0$.
After measuring the numerator and the denominator separately, one can obtain the value of Eq.~\eqref{eq:P2/I2} for given $r$ and $\tau$ with the precision $\epsilon_1+\tau\epsilon_2+\epsilon_3+\epsilon_\Lambda$.

\subsubsection{Three-point function \label{sec:3ptFunc}}
\noindent
Next, we discuss the three-point functions of the velocity field.
To reduce the computational complexity of reading out classical information, we introduce sub-grids, $x_{{\rm cg} k}\coloneq 2^{n_x-m}k\Delta x$, where an integer $m\geq1$ is taken independent of $n_x$, and coarse-grain the high-resolution $\vert\boldsymbol u\rangle$ onto the sub-grids.
This is motivated when the small-scale modes are nonsignificant because increasing $n_x$ beyond the smallest among the relevant scales of turbulence would not provide better information.
We can integrate out the small-scale modes within each sub-grid by acting Hadamard gates on the subspace corresponding to the last $n_x-m$ digits of each $\vert j\rangle$:
\begin{align}
    \vert\boldsymbol u\rangle_{\rm cg}
        &\coloneq1_{2^m}\otimes H^{\otimes n_x-m}\vert\boldsymbol u\rangle\notag\\
        &=\sum_{k=0}^{2^m-1}u_{{\rm cg}\, k}\vert k\rangle\vert 0\rangle_{\rm ss}+\vert\!\!\perp\rangle,
    \label{eq:cg}
\end{align}
where $H$ is the Hadamard gate,
\begin{align}
    u_{{\rm cg}\, k}
        \coloneq2^{-\frac{n_x-m}{2}}\sum_{j=2^{n_x-m}k}^{2^{n_x-m}(k+1)-1}\dfrac{u_j}{\vert\vert\boldsymbol u\vert\vert},
    \label{eq:ucgdef}
\end{align}
$\ket{0}_\mathrm{ss} \coloneq \ket{0}^{\otimes n_x-m}$, $1_{2^m}$ is the indentity opeartor on the first $m$ qubits, and $\ket{\perp}$ is a garbage state such that $\left(1_{2^m} \otimes \ket{0}_\mathrm{ss} \bra{0}_\mathrm{ss}\right)\ket{\perp}=0$.
Hereafter, we define $O_{u_{\rm cg}}\coloneq (1_{2^m}\otimes H^{\otimes n_x-m})O_{u_\tau}$ so that $O_{u_{\rm cg}}\vert\boldsymbol0\rangle=\vert\boldsymbol u\rangle_{\rm cg}$.

Then, a three-point function of the velocity field $P^{(3)}(r_1=2^{n_x-m}\rho_1\Delta x,r_2=^{n_x-m}\rho_2\Delta x)$, $\rho_1,\rho_2\in{\mathbb Z}_{2^m}$ can be approximated as
\begin{align}
    P^{(3)}\left(r_1,r_2\right)
        &\notag\\
        &\hspace{-18.5mm}
        \simeq2^{-n_x}\sum_{j=0}^{2^{n_x}-1}u_ju_{j+2^{n_x-m}\rho_1}u_{j+2^{n_x-m}\rho_2}\notag\\
        &\hspace{-18.5mm}
        \simeq2^{-\frac{3}{2}n_x+\frac{1}{2}m}\vert\vert\boldsymbol u\vert\vert^3\sum_{j,k,l=0}^{2^{m}-1}C^{(3)}_{jkl}(\rho_1,\rho_2)u_{{\rm cg}\,j}u_{{\rm cg}\,k}u_{{\rm cg}\,l}.
        \label{eq:3pt_matrixform}
\end{align}
Here, in the second equality, we replace $u_j$, $j\in[2^{n_x-m}k,2^{n_x-m}(k+1)-1]$ with
\begin{align}
    2^{-(n_x-m)}\sum_{j'=2^{n_x-m}k}^{2^{n_x-m}(k+1)-1}u_{j'}=2^{-\frac{n_x-m}{2}}\vert\vert\boldsymbol u\vert\vert u_{{\rm cg}\,k},
\end{align}
assuming that $u$ is coherent over the length scale $2^{-m}L$, {\it i.e.}, smaller-scale modes within the scale $2^{-m}L$ are subdominant.
We define
\begin{align}\hspace{-2mm}
    C^{(3)}_{jkl}(\rho_1,\rho_2)
        \coloneq\!\!\!\sum_{\sigma,\sigma',\sigma''}\!\!\!\delta_{j-\sigma,k-\sigma'}\delta_{j-\sigma,l-\sigma''}c^{(3)}_{\sigma\sigma'\sigma''}(\rho_1,\rho_2)
\end{align}
and
\begin{align}\hspace{-2mm}
    c^{(3)}_{\sigma\sigma'\sigma''}(\rho_1,\rho_2)
        \coloneq\mathcal S\left[\begin{cases}
            1&(\sigma,\sigma',\sigma'')=(0,\rho_1,\rho_2)\\
            0&\text{otherwise}
        \end{cases}\right],
\end{align}
where $\mathcal S$ is symmetrization with respect to the indices.
Note that the prefactor in front of the summation in Eq.~\eqref{eq:3pt_matrixform} scales as $\mathcal O(2^{m/2})$, independently of $n_x$.

From Eq.~\eqref{eq:3pt_matrixform}, we obtain an expression for the three-point function $P^{(3)}(r_1,r_2,\tau)$ normalized by the third moment $I^{(3)}(\tau)=P^{(3)}(0,0,\tau)$,
\begin{align}
    \dfrac{P^{(3)}\left(r_1,r_2,\tau\right)}{I^{(3)}(\tau)}
        =\dfrac{\left\langle U^{(3)}(\tau)\right\vert\tilde{C}^{(3)}(\rho_1,\rho_2)\left\vert U^{(3)}(\tau)\right\rangle}{\left\langle U^{(3)}(\tau)\right\vert\tilde{C}^{(3)}(0,0)\left\vert U^{(3)}(\tau)\right\rangle},
    \label{eq:P3/I3}
\end{align}
where we define
\begin{align}\hspace{-3mm}
    \bigl\vert U^{(3)}\bigr\rangle
        &\coloneq\dfrac{\left\vert{\boldsymbol{u}}\right\rangle_{\rm cg}^{\otimes2}\otimes\left\vert0\right\rangle+\left\vert{\boldsymbol{u}}\right\rangle_{\rm cg}\otimes\left\vert\boldsymbol{0}\right\rangle\otimes\left\vert1\right\rangle}{\sqrt{2}}\\
        &=\sum_{j,k}\dfrac{u_{{\rm cg}\,j}u_{{\rm cg}\,k}}{\sqrt{2}}\vert j,k,0\rangle+\sum_{l}\dfrac{u_{{\rm cg}\,l}}{\sqrt{2}}\vert l,0,1\rangle+\ket{\perp},
\end{align}
and
\begin{align}
    \tilde{C}^{(3)}
        \coloneq\sum_{j,k,l=1}^{N_x}C^{(3)}_{jkl}\dfrac{\vert j,k,0\rangle\langle l,0,1\vert+\vert l,0,1\rangle\langle j,k,0\vert}{2}\notag\\
        \otimes\vert 0\rangle_{\rm ss}\langle0\vert_{\rm ss},
\end{align}
where $\ket\perp$ is the garbage state originating from the one that $\left\vert{\boldsymbol{u}}\right\rangle_{\rm cg}$ has, and the projection $\vert 0\rangle_{\rm ss}\langle0\vert_{\rm ss}$ eliminates the garbage state.
Note that, since the prefactor in the last line in Eq.~\eqref{eq:3pt_matrixform} scales as $2^{m/2}$, the numerator and the denominator in the right-hand side of Eq.~\eqref{eq:P3/I3} are $\mathcal O(2^{-m/2})$.

The state $\bigl\vert U^{(3)}\bigr\rangle=O_{U^{(3)}}\vert0\rangle$, where $O_{U^{(3)}}\coloneq O_{u_{\rm cg}}\otimes\left[CO_{u_{\rm cg}}(1_{N_x}\otimes H)\right]$, and $H$ is the Hadamard gate, can be constructed if we have $CO_{u_{\rm cg}} \coloneq 1_{N_x} \otimes \ket{0}\bra{0}+O_{u_{\rm cg}} \otimes \ket{1}\bra{1}$, namely a controlled-$O_{u_{\rm cg}}$ operation. 
In the complexity estimate, we assume that $CO_{u_{\rm cg}}$ requires the same query complexity as $O_{u_{\rm cg}}$ does.

The operator $\tilde{C}^{(3)}$ is a matrix of ${\mathcal O}(1)$ sparsity and maximum norm, and it has an $({\mathcal O}(1),{\mathcal O}(n_x),\epsilon_4)$ block-encoding $\mathcal{C}^{(3)}$ with ${\mathcal O}(n_x)$ ancilla qubits.
To construct this, we make $\mathcal{O}(1)$ queries to the sparse-access oracles for $\tilde{C}^{(3)}$ and use
\begin{align}
    {\mathcal O}\left(n_x+\log^{\frac{5}{2}}\left(\dfrac{1}{\epsilon_4}\right)\right)\;{\text{\it fundamental gates}}
    \label{eq:BE_complexity}
\end{align}
and $\mathcal{O}(\log^{5/2}(1/\epsilon_4))$ ancilla qubits \cite{gilyen2019quantum}.

Since $\mathcal{C}^{(3)}$ is unitary, the numerator and the denominator in the right-hand side of Eq.~\eqref{eq:P3/I3} can be measured with the error tolerance $2^{-m/2}\epsilon_3$ by the overlap estimation algorithm \cite{2007PhRvA..75a2328K}, which makes
\begin{align}
    {\mathcal{O}}\left(\dfrac{2^{\frac{m}{2}}}{\epsilon_3}\right)\;\text{\it queries to}\;{\mathcal C}^{(3)} \;\text{\it and}\;{O}_{U^{(3)}}.
    \label{eq:OEEfor3pt}
\end{align}
Then, after measuring the numerator and the denominator separately, one may obtain the value of Eq.~\eqref{eq:P3/I3} for given $r_1,$ $r_2$, and $t$, with an error tolerance $\epsilon_3$.
Note that, if one skips the coarse-graining procedure by taking $m=n_x$, we need $2^{-n_x/2}\epsilon_3$ accuracy in the overlap estimation algorithm to achieve an error tolerance $\epsilon_3$ in evaluating  Eq.~\eqref{eq:P3/I3}, implying an exponentially larger query complexity in Eq.~\eqref{eq:OEEfor3pt}.

\subsubsection{Higher-order multi-point function}
\noindent
We then generalize the procedure to the higher-order multi-point functions.
For a general integer $n$, we may approximate
\begin{align}
    P^{(n)}({\boldsymbol r})
        &\notag\\
        &\hspace{-8mm}=2^{-\frac{n}{2}n_x+\frac{n-2}{2}m}\vert\vert\boldsymbol u\vert\vert^n\sum_{\boldsymbol j}C^{(n)}_{\boldsymbol j}(\boldsymbol\rho)\prod_{i=1}^n u_{{\rm cg}\,j_i},
        \label{eq:npt_matrixform}
\end{align}
where $\boldsymbol{r}=(r_1,\cdots, r_{n-1})$, $r_0=0,$ $\boldsymbol{\rho}=2^{-n_x+m}\boldsymbol{r}/\Delta x\in\mathbb Z^{n-1}$, and $\boldsymbol{j}=(j_1,\cdots j_n)$, respectively. 
The matrix $C^{(n)}$ is defined such that
\begin{align}
    C^{(n)}_{\boldsymbol j}\left({\boldsymbol\rho}\right)
        &\coloneq\sum_{\boldsymbol{\sigma}}\delta_{\boldsymbol j-\boldsymbol\sigma}c^{(n)}_{\boldsymbol\sigma}\left({\boldsymbol\rho}\right),\\
    c^{(n)}_{\boldsymbol\sigma}\left({\boldsymbol\rho}\right)
        &\coloneq{\mathcal S}\left[\left\{\begin{matrix}
        1&{\boldsymbol\sigma}={\boldsymbol\rho},\\[7pt]
        0&\text{otherwise}
        \end{matrix}\right.\right],
\end{align}
where ${\mathcal S}$ is symmterization with respect to the indices $\sigma_i$.

If $n=2n'$ is an even number, we define
\begin{align}
    \bigl\vert U^{(2n')}\bigr\rangle
        &\coloneq{O}_{U^{(2n')}}\vert0\rangle
        =\left\vert{\boldsymbol u}\right\rangle_{\rm cg}^{\otimes n'},
        \label{eq:U2n}
\end{align}
where ${O}_{U^{(2n')}}\coloneq{O}_{u_{\rm cg}}^{\otimes n'}$ is constructed by the following quantum circuit,

\hspace{5mm}\begin{minipage}[t]{4cm}\vspace{2mm}
\begin{quantikz}
\lstick{$\ket{0}\;\,$}
\gategroup[2,steps=7,style={dashed,rounded corners,inner xsep=2pt,xshift=-0.8cm},label style={label position=above,anchor=south,yshift=-0.2cm}]{
$\mathcal{O}_{u_{\rm cg}}$ (Secs.~\ref{sec:C2Q} and \ref{sec:Q})
}
&\qwbundle{n_x}&\gate{\mathcal{O}_{\psi'_0}}&\gate[2]{e^{A\tau}}&\gate[2]{\rm cg}&&\rstick[5]{$\ket{U^{(2n')}}$} \\
\lstick{$\ket{0}_a$}&\qwbundle{}&&&&&\\[7pt]
\wave&&&&&&\\[7pt]
\lstick{$\ket{0}\;\,$}
\gategroup[2,steps=7,style={dashed,rounded corners,inner xsep=2pt,xshift=-0.8cm},label style={label position=above,anchor=south,yshift=-0.2cm}]{}
&\qwbundle{n_x}&\gate{\mathcal{O}_{\psi'_0}}&\gate[2]{e^{A\tau}}&\gate[2]{\rm cg}&&\\
\lstick{$\ket{0}_a$}&\qwbundle{}&&&&&
\end{quantikz}\vspace{2mm}
\end{minipage}

\noindent
where cg denotes the coarse-graining procedure.
We also define
\begin{align}
    \tilde{C}^{(2n')}
        \coloneq\sum_{\boldsymbol{j}}&C^{(2n')}_{\boldsymbol{j}}\biggl(\dfrac{\vert j_1,\cdots, j_{n'}\rangle\langle j_{n'+1},\cdots, j_{2n'}\vert}{2}\notag\\
        &\quad+\dfrac{\vert j_{n'+1},\cdots, j_{2n'}\rangle\langle j_1,\cdots, j_n'\vert}{2}\biggr).
\end{align}
This is $n!$-sparse and has an $({\mathcal O}(n!),{\mathcal O}(nn_x),\epsilon_4)$ block-encoding $\mathcal C^{(n)}$, which is constructed with 
\begin{align}
    \mathcal{O}(1)~{\text{\it queries to the sparse-access oracles for }} \tilde{C}^{(n)}
    \label{eq:CnComp}
\end{align}
and
\begin{align}
    {\mathcal O}\left(nn_x+\log^{\frac{5}{2}}\left(\dfrac{n!}{\epsilon_4}\right)\right)\;{\text{\it fundamental gates}}.
    \label{eq:CnGate}
\end{align}

If $n=2n'+1$ is an odd number, we define
\begin{align}
    \bigl\vert U^{(2n'+1)}\bigr\rangle
        &\coloneq O_{U^{(2n'+1)}}\vert0\rangle\notag\\
        &=\dfrac{\left\vert{\boldsymbol{u}}\right\rangle_{\rm cg}^{\otimes n'+1}\otimes\vert0\rangle+\left\vert{\boldsymbol{u}}\right\rangle_{\rm cg}^{\otimes n'}\otimes\vert{\boldsymbol 0}\rangle\otimes\vert1\rangle}{\sqrt{2}},
        \label{eq:U2n+1}
\end{align}
where $O_{U^{(2n'+1)}}\coloneq {O}_{u_{\rm cg}}^{\otimes n'}\otimes\left[C{O}_{u_{\rm cg}}(1_{N_x}\otimes H)\right]$ is constructed by the following quantum circuit,

\hspace{-3mm}\begin{minipage}[t]{4cm}\vspace{2mm}
\begin{quantikz}
\lstick{$\ket{0}\;\,$}
\gategroup[2,steps=7,style={dashed,rounded corners,inner xsep=2pt,xshift=-0.8cm},label style={label position=above,anchor=south,yshift=-0.2cm}]{
$\mathcal{O}_{u_{\rm cg}}$ (Secs.~\ref{sec:C2Q} and \ref{sec:Q})
}
&\qwbundle{n_x}&\gate{\mathcal{O}_{\psi'_0}}&\gate[2]{e^{A\tau}}&\gate[2]{\rm cg}&&\rstick[6]{$\ket{U^{(2n'+1)}}$} \\
\lstick{$\ket{0}_a$}&\qwbundle{}&&&&&\\[7pt]
\wave&&&&&&\\[7pt]
\lstick{$\ket{0}\;\,$}
\gategroup[2,steps=7,style={dashed,rounded corners,inner xsep=2pt,xshift=-0.8cm},label style={label position=above,anchor=south,yshift=-0.2cm}]{}
&\qwbundle{n_x}&\gate{\mathcal{O}_{\psi'_0}}&\gate[2]{e^{A\tau}}&\gate[2]{\rm cg}&&\\
\lstick{$\ket{0}_a$}&\qwbundle{}&&&&&\\
\lstick{$\ket{0}\;\,$}&\gate{H}&\ctrl[open]{-2}&\ctrl[open]{-1}&\ctrl[open]{-1}&&
\end{quantikz}\vspace{2mm}
\end{minipage}

\noindent
and
\begin{align}
    &\hspace{-3mm}\tilde{C}^{(2m+1)}\notag\\
        \coloneq&\sum_{\boldsymbol{j}}C^{(2n'+1)}_{\boldsymbol{j}}\biggl(\dfrac{\vert j_1,\cdots, j_{n'+1},0\rangle\langle j_{n'+2},\cdots, j_{2n'+1},0,1\vert}{2}\notag\\
        &\quad\;\;+\dfrac{\vert j_{n'+2},\cdots, j_{2n'+1},0,1\rangle\langle j_1,\cdots, j_{n'+1},0\vert}{2}\biggr),
\end{align}
which also has an $({\mathcal O}(n!),{\mathcal O}(nn_x),\epsilon_4)$ block-encoding with the same cost as Eqs.~\eqref{eq:CnComp} and \eqref{eq:CnGate}.

Similarly to the case of $P^{(3)}$, we can obtain
\begin{align}
    \dfrac{P^{(n)}({\boldsymbol r},\tau)}{I^{(n)}(\tau)}
        =\dfrac{\left\langle U^{(n)}(\tau)\right\vert\tilde{C}^{(n)}({\boldsymbol\rho})\left\vert U^{(n)}(\tau)\right\rangle}{\left\langle U^{(n)}(\tau)\right\vert\tilde{C}^{(n)}({\boldsymbol 0})\left\vert U^{(n)}(\tau)\right\rangle}
    \label{eq:npt}
\end{align}
by applying the overlap estimation algorithm \cite{2007PhRvA..75a2328K}.
To achieve error tolerance $\epsilon_3$, we have to make
\begin{align}
    {\mathcal{O}}\left(\dfrac{2^{\frac{n-2}{2}m}}{\epsilon_3}\right)\;\text{\it queries to}\;{\mathcal C}^{(n)} \;\text{\it and}\;{O}_{U^{(n)}}.
    \label{eq:OEEfornpt}
\end{align}

\subsection{Taking ensemble averages\label{sec:En}}
\noindent
So far, we have discussed computing spatial correlations of the velocity field for a single realization of Burgers turbulence.
As discussed in Sec.~D.2 of Ref.~\cite{Miyamoto:2023iwk}, we can extend it to taking ensemble averages of multiple realizations.

To this end, we assume that we have $N_{\rm en}=2^{n_{\rm en}}$ initial conditions and have an oracle 
\begin{align}
    O_{\rm IC}
        \coloneq \left(\sum_{\alpha=1}^{N_{\rm en}}O_{\psi'^{(\alpha)}_0}\otimes\left\vert\alpha\right\rangle\left\langle \alpha\right\vert\right)\left(1_{N_x} \otimes H^{\otimes n_{\rm en}}\right)
\end{align}
so that a superposition of $N_{\rm en}$ initial conditions are generated as
\begin{align}
    O_{\rm IC}\left\vert0\right\rangle
        =2^{-\frac{n_{\rm en}}{2}}\sum_{\alpha=1}^{N_{\rm en}}\left\vert\partial_x\boldsymbol{\psi}^{(\alpha)}(0)\right\rangle\left\vert\alpha\right\rangle,
    \label{eq:superposedIC}
\end{align}
where we introduce additional $n_{\rm en}$ qubits to label the ensembles.

We then integrate the heat equation as explained in Sec.~\ref{sec:Q}.
We apply ${O}_{\psi'_\tau}\otimes 1_{N_{\rm en}}$ to the state \eqref{eq:superposedIC} with error tolerance $\epsilon_1$ for the integration algorithm and $\epsilon_2$ for the block-encoding of $A$.
With the approximation \eqref{eq:approx_u_partialpsi}, this yields
\begin{align}
    2^{-\frac{n_{\rm en}}{2}}\sum_{\alpha=1}^{N_{\rm en}} \ket{\boldsymbol{u}^{(\alpha)}}\ket{\alpha},
\end{align}
where $\ket{\boldsymbol{u}^{(\alpha)}}:=\sum_j\dfrac{u^{(\alpha)}_j}{\vert\vert\boldsymbol{u}^{(\alpha)}\vert\vert}\vert j\rangle$ encodes the solution $u^{(\alpha)}$ with the $\alpha$-th initial condition.
By applying the coarse-graining operation described in Sec. \ref{sec:3ptFunc}, we obtain
\begin{align}
    2^{-\frac{n_{\rm en}}{2}}\sum_{\alpha=1}^{N_{\rm en}} \ket{\boldsymbol{u}^{(\alpha)}}_\mathrm{cg}\ket{\alpha},
\end{align}
where $\ket{\boldsymbol{u}^{(\alpha)}}_\mathrm{cg}$ is given as Eq.~\eqref{eq:cg} with $u=u^{(\alpha)}$.

Then, using the above operation, we can generate
\begin{align}
    \ket{U^{(n)}}_\mathrm{en} \coloneq 2^{-\frac{n_{\rm en}}{2}}\sum_{\alpha=1}^{N_\mathrm{en}}\vert U^{(n, \alpha)}\rangle\vert\alpha\rangle, 
\end{align}
where $U^{(n, \alpha)}$ is given as Eq.~\eqref{eq:U2n} or \eqref{eq:U2n+1} with $u=u^{(\alpha)}$.
Defining the ensemble average of the $n$-point function as
\begin{align}
    \left\langle P^{(n)}({\boldsymbol r})\right\rangle
        &\coloneq2^{-n_{\rm en}}\sum_{\alpha=1}^{N_{\rm en}}P^{(n,\alpha)}({\boldsymbol r}),
    \label{eq:Pn_en}
\end{align}
where $P^{(n,\alpha)}({\boldsymbol r})$ is the $n$-point function for $u^{(\alpha)}$, and $\left\langle I^{(n)} \right\rangle \coloneq \left\langle P^{(n)}({\boldsymbol 0})\right\rangle$, we see that
\begin{align}
    \frac{\left\langle P^{(n)}({\boldsymbol r})\right\rangle}{\left\langle I^{(n)} \right\rangle} = \frac{\bra{U^{(n)}}_\mathrm{en}\tilde C^{(n)}(\boldsymbol\rho)\otimes1_{N_\mathrm{en}}\ket{U^{(n)}}_\mathrm{en}}{\bra{U^{(n)}}_\mathrm{en}\tilde C^{(n)}(\boldsymbol 0)\otimes1_{N_\mathrm{en}}\ket{U^{(n)}}_\mathrm{en}},
\end{align}
and thus we can calculate this in a parallel way to the procedure in Sec.~\ref{sec:Q2C}.
Namely, we perform the overlap estimation with an error tolerance $\epsilon_3$, by replacing $C^{(2)}$ and $\mathcal C^{(n)}$ with $C^{(2)}\otimes 1$ and $\mathcal C^{(n)}\otimes 1$, respectively.

\section{Discussion\label{sec:Discussion}}
\subsection{Total complexity}
\noindent
We first estimate the complexity of the algorithm, apart from the discretization error, which is common in both quantum and classical algorithms and may be reduced by implementing higher-order stencils.
Our algorithm accepts errors of ${\mathcal O}(\epsilon_1+\tau\epsilon_2+\epsilon_3+\epsilon_4+\epsilon_\Lambda)$ for $P^{(n)}/I^{(n)}$. 
By taking $\tau\epsilon_2\sim\epsilon$ and $\epsilon_1\sim\epsilon_3\sim\epsilon_4\sim\epsilon_\Lambda\sim\epsilon$ to be of the order of the desired error tolerance $\epsilon$, we require query complexities at each step of the operation as summarized in TABLE \ref{tb:complexity}.
In total, to measure $P^{(n)}/I^{(n)}$ for a general $n$, we make
\begin{align}\hspace{-3mm}
    &\mathcal{O}\left(2^{\frac{n-2}{2}m}n\dfrac{\vert\vert\partial_x\boldsymbol{\psi}_0\vert\vert}{\vert\vert\partial_x\boldsymbol{\psi}(\tau)\vert\vert}\dfrac{\tau}{\epsilon}{\rm polylog}{\left(\dfrac{1}{\epsilon}\right)\left(n_x+\log^{\frac{5}{2}}\left(\dfrac{\tau}{\epsilon}\right)\right)}\right)
    \notag\\
        =&\tilde{\mathcal{O}}\left(2^{\frac{n-2}{2}m}n\dfrac{\vert\vert\partial_x\boldsymbol{\psi}_0\vert\vert}{\vert\vert\partial_x\boldsymbol{\psi}(\tau)\vert\vert}\dfrac{n_x\tau}{\epsilon}\right)
    \label{eq:complexity}
\end{align}
uses of sparse-access queries and other fundamental gates and use
\begin{align}
    \mathcal{O}\left(n_x+\log^{\frac{5}{2}}\left(\dfrac{\tau}{\epsilon}\right)\right)
\end{align}
ancilla qubits, where $n_x=\log N_x$ and $\tau=\nu t/\Delta x^2$.
In particular, for $n=2$, we can use Eq.~\eqref{eq:C2_decomp} to construct $\mathcal{C}^{(n=2)}$ with $\mathcal O(n_x^2)$ fundamental gates instead of applying the sparse-oracle-based block-encoding technique, and the number of uses of other oracles and fundamental gates is also reduced.
Note that, if $\partial\psi_0$ has large long-wavelength modes, which dissipate slowly, then the growth of the factor $\vert\vert\partial_x\boldsymbol{\psi}_0\vert\vert/\vert\vert\partial_x\boldsymbol{\psi}(\tau)\vert\vert$ is slow as well.

This can be significantly less than a naive estimate of the classical complexity.
Since $\psi(x,t)$ is the convolution of $\psi(0,t)$ and the heat kernel, it takes ${\mathcal O}(N_x)$ computations to obtain $u(x,t)$ for a given $(x,t)$.
To compute $P^{(n)}$, we identify the ensemble average with the spatial average and compute $u(x,t)$ at all sites $x_j$ for a given $t$, which costs ${\mathcal O}(N^2_x)$ computations in total.
Therefore, when one increases $N_x$ to achieve high-resolution simulation, our quantum algorithm is exponentially faster.

\begin{table*}[htbp]
  \centering
  \begin{tabular}{c|c|c|cc}
  purpose & $n$ & oracle & number of uses & \\
  \hline \hline
  \makecell{initial value \\ encoding} & - & $O_{\psi'_0}$ & $\mathcal{O}\left(\dfrac{\vert\vert\partial_x\boldsymbol{\psi}_0\vert\vert}{\vert\vert\partial_x\boldsymbol{\psi}(\tau)\vert\vert}\dfrac{2^{\frac{n-2}{2}m}n}{\epsilon}\right)$ & = $n\times$Eq.~\eqref{eq:TEquery2}$\times$Eq.~\eqref{eq:OEEfornpt} \\
  \hline
  \multirow{2}{*}{time integration} & - & \makecell{sparse-access \\ oracles for $A$} & $\mathcal{O}\left(\dfrac{\vert\vert\partial_x\boldsymbol{\psi}_0\vert\vert}{\vert\vert\partial_x\boldsymbol{\psi}(\tau)\vert\vert}\dfrac{2^{\frac{n-2}{2}m}n\tau}{\epsilon} \mathrm{polylog}\left(\dfrac{1}{\epsilon}\right)\right)$ & $=n\times$Eq.~\eqref{eq:TEquery1}$\times$Eq.~\eqref{eq:BEquery2}$\times$Eq.~\eqref{eq:OEEfornpt} \\
  & - & \makecell{other \\ fundamental \\ gates} & $\mathcal{O}\left(\dfrac{\vert\vert\partial_x\boldsymbol{\psi}_0\vert\vert}{\vert\vert\partial_x\boldsymbol{\psi}(\tau)\vert\vert}\dfrac{2^{\frac{n-2}{2}m}n\tau}{\epsilon}\left(n_x+\log^{\frac{5}{2}}\left(\dfrac{\tau}{\epsilon}\right)\right) \mathrm{polylog}\left(\dfrac{1}{\epsilon}\right)\right)$ & $=n\times$Eq.~\eqref{eq:TEquery1}$\times$Eq.~\eqref{eq:BEquery1}$\times$Eq.~\eqref{eq:OEEfornpt} \\
  \hline
  \multirow{3}{*}{\makecell{block-encoding \\ of $\tilde{C}^{(n)}$}} & \multirow{2}{*}{$n\ge3$} & \makecell{sparse-access \\ oracles for $\tilde{C}^{(n)}$} & $\mathcal{O}\left(\dfrac{2^{\frac{n-2}{2}m}}{\epsilon}\right)$ & $=$Eq.~\eqref{eq:CnComp}$\times$ Eq.~\eqref{eq:OEEfornpt} \\
  & & \makecell{other \\ fundamental \\ gates} &  $\mathcal{O}\left(\dfrac{2^{\frac{n-2}{2}m}}{\epsilon}\left(nn_x+\log^{\frac{5}{2}}\left(\dfrac{n!}{\epsilon}\right)\right)\right)$ & $=$Eq.~\eqref{eq:CnGate}$\times$ Eq.~\eqref{eq:OEEfornpt} \\
  \cline{2-5}
  & 2 & \makecell{fundamental \\ gates} & $\mathcal{O}\left(\dfrac{n_x^2}{\epsilon}\right)$ & $=$Eq.~\eqref{eq:C2GateCost}$\times$Eq.~\eqref{eq:OEEfornpt}
  \end{tabular}
  \caption{The number of uses of various oracles in the proposed method to calculate the $n$-point function. Subdominant contributions and additional ones for controlled operations are neglected. Note that the complexity for block-encoding $\Lambda$ is always subdominant and omitted in the table.}
  \label{tb:complexity}
\end{table*}

\subsection{Validity of the assumptions and generalization}
\noindent
Next, let us clarify what assumptions are made for our algorithm to work and discuss possible generalizations to go beyond the direct application of the algorithm.

\subsubsection{Periodic boundary condition}
\noindent
In Sec.~\ref{sec:Q}, we imposed the periodic boundary condition on $\partial_x\psi$, namely $\partial_x\psi_{-1}=\partial_x\psi_{N_x-1}$ and $\partial_x\psi_{N_x}=\partial_x\psi_{0}$.
We can straightforwardly generalize the algorithm to another boundary condition, say the Dirichlet boundary condition, by just modifying the matrix $A$ and $C^{(n)}$.
The complexity estimate of our algorithm does not change with this modification.

One may further assume the periodic boundary condition in a stronger sense, namely $\psi_{-1}=\psi_{N_x-1}$ and $\psi_{N_x}=\psi_{0}$. Indeed, we assume that this condition holds when we take $\tilde\psi_j=\bar\psi$ as the zero-th order solution (see the following subsections).
In this case, as understood from the definition of $\psi$, Eq.~\eqref{eq:CHtrsf}, we implicitly limit ourselves to the situations where the global momentum of the system vanishes, $\int_0^Lu(x)dx=0$.

Unfortunately, generalizing the discussion to a globally moving system, {\it i.e.}, $\int_0^Lu(x)dx\neq0$, is not straightforward. Assuming the periodic boundary conditions on $u$ and $\partial_x u$, we may define
\begin{align}
    \psi_{-1}&=\dfrac{\psi_0\psi_{N_x-2}}{\psi_{N_x-1}},&
    \psi_{N_x}&=\dfrac{\psi_1\psi_{N_x-1}}{\psi_{0}},\notag\\
    \partial_x\psi_{-1}&=\dfrac{\psi_0(\partial_x\psi_{N_x-2})}{\psi_{N_x-1}},&
    \partial_x\psi_{N_x}&=\dfrac{(\partial_x\psi_1)\psi_{N_x-1}}{\psi_{0}},
\end{align}
where we use the central differencing scheme.
Apparently, we cannot incorporate these nonlinear expressions into the matrices $A$ and $C^{(n)}$.

Nevertheless, we can address such a situation by taking a sufficiently large box and $u(x\sim0)=u(x\sim L)=0$ throughout the evolution.
In that case, we do not have to modify the matrices $A$ and $C^{(n)}$ because
\begin{align}
    \partial_x\psi_{-1}=\partial_x\psi_{N_x}=\partial_x\psi_{j\sim0}=\partial_x\psi_{j\sim N_x-1}=0.
\end{align}

\subsubsection{Approximation in Eq.~\eqref{eq:dpsi_to_u}}
\noindent

Now, let us consider the applicability of Eq.~\eqref{eq:dpsi_to_u}.
First, we consider decomposing $\psi$ into the homogeneous and fluctuation parts as $\psi(x)=\bar\psi+\delta\psi(x)$ and taking $\tilde{\psi}=\bar\psi$.
This approximation is not always valid for a large Reynolds number: a naive estimate $\delta\psi/\bar\psi\sim{\rm Re}$ implies that the equation is not a good approximation when the Reynolds number is large.
That is, although the state $\vert\partial_{\boldsymbol x}\boldsymbol{\psi}(t)\rangle$ always ``knows" the solution $u(t,x)$ in a sense, we can efficiently measure the informative quantities, $P^{(n)}/I^{(n)}$, only when the Reynolds number is small enough.

\begin{figure}[ht]
        \begin{minipage}[h]{1.0\hsize}
            \includegraphics[keepaspectratio, width=0.96\textwidth]{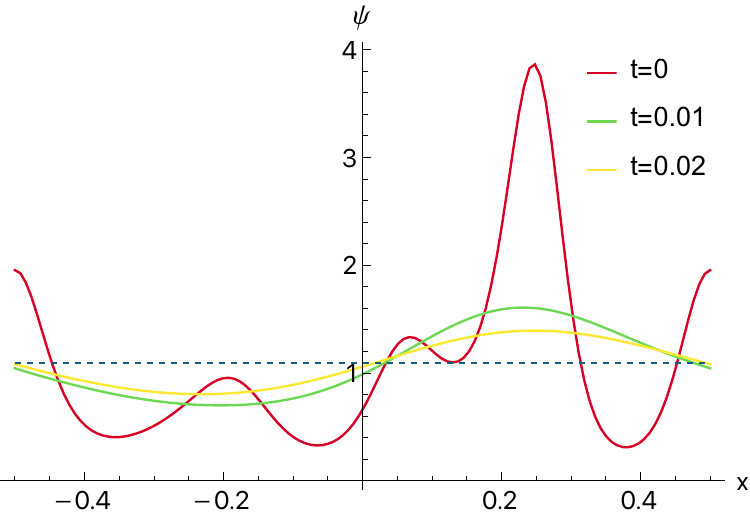}
        \end{minipage}
    \vspace{1mm}
    \caption{\label{fig:psi}An example of the evolution of $\psi$ with a random initial condition given in Eq.~\eqref{eq:randomini} (Red-solid line at $t=0$, green-solid at $t=0.01$, and yellow-solid at $t=0.02$). We also show the spatial average $\bar\psi$, which is constant in time,} with the blue-dashed line. Our numerical code used to make this plot is publically available at the URL given in Ref. \cite{mycode}.
\end{figure}

\begin{figure}[ht]
        \begin{minipage}[h]{1.0\hsize}
            \includegraphics[keepaspectratio, width=0.96\textwidth]{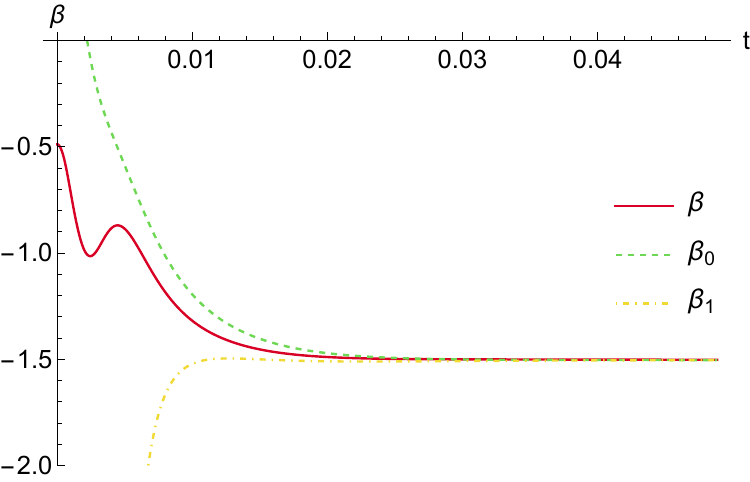}
        \end{minipage}
    \vspace{1mm}
    \caption{\label{fig:beta}Comparison between the values of flatness $\beta$ obtained by the zero-th order approximation, Eq.~\eqref{eq:dpsi_to_u} with $\tilde\psi_j=\bar\psi$} (green-dashed), the first-order approximation, \eqref{eq:approx_1st} (yellow dash-dot), and the exact relation Eq.~\eqref{eq:psi_to_u} (red-solid). Our numerical code used to make this plot is public at \cite{mycode}.
\end{figure}

\begin{figure}[ht]
        \begin{minipage}[h]{1.0\hsize}
            \includegraphics[keepaspectratio, width=0.96\textwidth]{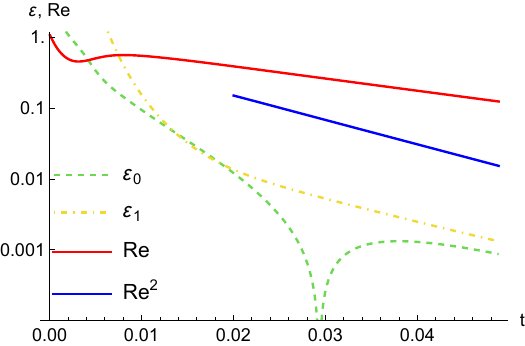}
        \end{minipage}
    \vspace{1mm}
    \caption{\label{fig:Re}The evolution of the Reynolds number (red-solid) and of its square (blue-solid), and the relative errors of the flatness, $\varepsilon_0:=\vert\beta/\beta_0-1\vert$ (green-dashed) and $\varepsilon_1:=\vert\beta/\beta_1-1\vert$ (yellow dash-dot). 
    Our numerical code used to make this plot is public at \cite{mycode}.}
\end{figure}

To classically demonstrate the difference between using the approximation Eq.~\eqref{eq:dpsi_to_u} and the exact relation Eq.~\eqref{eq:psi_to_u}, we take a random initial condition,
\begin{align}
    \psi(x,0)
        =e^{\xi_0+\sum_{j=1}^{j_{\rm max}}\left(\xi_j\cos\frac{2\pi jx}{L}+\xi_{j_{\rm max}+j}\sin\frac{2\pi jx}{L}\right)},
    \label{eq:randomini}
\end{align}
where $\xi_j$, $j=1,\cdots, 2j_{\rm max}$ are random variables generated by a normal distribution with the expectation value $0$ and the standard deviation $\sigma_\xi$.
Here we take $\sigma_\xi=0.3$, $j_{\rm max}=5$, and $n_x=7$, and obtain an initial condition of $\psi$ as shown in the red-solid line in Fig.~\ref{fig:psi}.
The green-solid and yellow-solid lines represent the evolution of $\psi$, and the blue-dashed line represents $\bar\psi$, where we take $\nu=1$.
We see that a long-wavelength mode survives and $\psi$ typically deviates from $\bar\psi$ by more than $\sim10$\% until $t=0.02$.
Correspondingly, the Reynolds number, estimated as ${\rm Re}\simeq\nu^{-1}u_{\rm rms}^2/(\partial_xu)_{\rm rms}$, where the subscript rms indicates the root mean square of variables, is ${\rm Re}\simeq0.38$ at $t=0.02$ (see Fig.~\ref{fig:Re}).
Small Reynolds number ${\rm Re}\ll1$ implies the perturbative regime, where we expect that the relative error originated from the approximation in Eq.~\eqref{eq:dpsi_to_u} is suppressed by powers of the small factor $\mathcal O({\rm Re})$.
As a test of the validity of the approximation, we plot in Fig.~\ref{fig:beta} the flatness $\beta$ defined as
\begin{align}
    \beta
        \coloneq\dfrac{I^{(4)}}{\left(I^{(2)}\right)^2}-3.
        \label{eq:flatnessdef}
\end{align}
This generally approaches $\beta=-3/2$ in a perturbative regime (see Appendix \ref{appx:analytic}), and we thus expect that calculating $\beta$ would be useful as a test also in quantum computing on future real hardware.
In Fig.~\ref{fig:beta}, both the exact value (solid line) and the approximation (dashed line) converge to the asymptotic value in the perturbative regime, $\beta=-3/2$, at $t\simeq0.02$.
We see that our approximation works in the late dissipative regime, even though $\delta\psi$ is set large at the beginning, and also that even the zero-th order approximation $\beta_0$, obtained by Eq.~\eqref{eq:dpsi_to_u} with $\tilde\psi_j=\bar\psi$, gives an accurate approximation of $\beta$.
In Fig.~\ref{fig:Re}, we show the evolution of $\varepsilon_0:=\vert\beta/\beta_0-1\vert$ and $\varepsilon_1:=\vert\beta/\beta_1-1\vert$, the relative errors of the $\beta$ in the zero-th and first order approximations, along with the Reynolds number and its square. Note that, in computing $\beta$, zero-th order approximation is good enough, since the correction begins at the second order (see Appendix \ref{appx:analytic}), which is indicated by the behaviours of $\varepsilon_0$ and $\varepsilon_1$ follow the evolution of $\mathrm{Re}^2$ in the perturbative regime, $t\gtrsim0.02$.

We can even go beyond the zero-th order approximation, provided that the Reynolds number is small.
We decompose $\tilde\psi$ in the denominator of Eq.~\eqref{eq:dpsi_to_u} into $\tilde\psi=\psi^{(0)}+\delta\psi$, and then we have the first-order approximation
\begin{align}
    u_j
        \simeq-\dfrac{\nu}{\psi_j^{(0)}}\partial_x \psi_j
        +\dfrac{\nu}{\psi^{(0)2}_j}\delta\psi_j\partial_x \psi_j \simeq -\frac{\nu}{\psi^{(0)2}_j}\phi_j \partial_x \psi_,
    \label{eq:approx_1st}
\end{align}
where $\phi \coloneq \psi - 2\delta \psi \simeq \psi^{(0)} - \delta \psi$, and $\mathcal N''$ is a normalization factor.
We can obtain $\vert\boldsymbol\phi\rangle$ in the same manner as $\vert\partial_x\boldsymbol\psi\rangle$ because $\phi$ obeys the heat equation similar to Eq. \eqref{eq:d_heateqn},
and thus estimate $\overline{u(x)u(x+r)}$ via
\begin{align}
\overline{~ \frac{1}{\psi^{(0)}(x)\psi^{(0)}(x+r)}\phi(x)\,\partial_x\psi(x)\,\phi(x+r)\,\partial_x\psi(x+r) ~}
\end{align}
by an extension of the method for estimating four-point functions described in Sec.~\ref{sec:Q2C}, whose complexity is given as Eq. \eqref{eq:complexity} with $n=4$.
That is, we can obtain the two-point function of $u$ with first-order accuracy, but the complexity increases to that for the zeroth-order four-point function.
For reference, we plot our first-order approximation $\beta_1$, obtained by Eq.~\eqref{eq:approx_1st} with $\psi_j^{(0)}=\bar\psi$, in Figs.~\ref{fig:beta} and \ref{fig:Re}.

\subsubsection{Choice of the background approximation}
\noindent
We may choose a homogeneous background solution, $\tilde\psi=\bar\psi$, as long as $\psi$ satisfies the periodic boundary condition and the Reynolds number is small.
However, since not $\psi$ but $u$ is the physical observable, it would be reasonable to assume the periodic boundary condition on $u$.
In this case, $\psi$ does not satisfy the periodic boundary condition and does not necessarily have a homogeneous background.

We then propose to set another choice of the background solution for $\psi$.
First, observe that the Burgers equation conserves the momentum within a periodic box.
Therefore, we have an integral of motion, $\bar u\coloneq\frac{1}{L}\int_0^Lu(x)dx$.
We may approximate $u(x)$ by $\bar u$ as the zero-th order approximation\footnote{For instance, the Brownian motion, which is sometimes taken in the literature as the initial turbulent condition \cite{1992CMaPh.148..623S,1992CMaPh.148..601S,1998CMaPh.193..397B,2009JSP...134..589V}, is beyond its applicability. In this case, $\psi(x)$ scales not exponentially but as $\sim\exp(-\alpha x^{3/2})$, where $\alpha$ is some constant (in $x$). However, the system eventually enters the perturbative regime, where the approximation in Eq.~\eqref{eq:approx_bkgpsi}, holds.}, and then we obtain
\begin{align}
    \tilde\psi_j
        =\kappa(t)\exp\left(-\dfrac{\bar u \Delta x}{2\nu}j\right),
    \label{eq:approx_bkgpsi}
\end{align}
where $\kappa(t)$ is a $j$-independent growth factor.
If we neglect the contribution from inhomogeneous modes, $\kappa(t)\propto\exp(\bar u^2 t/4\nu)$.
Note that $\kappa(t)$ does not matter because of the normalization of the quantum state $\vert \partial_x\boldsymbol\psi\rangle$.
In particular, taking $\bar\psi$ as the background solution can be regarded as a special case where $\bar u=0$.

\section{Conclusion \label{sec:Conclusion}}
\noindent
We developed an FTQC algorithm to measure the statistical properties of Burgers turbulence.
By applying the Cole–Hopf transformation, we are able to effectively linearize the nonlinear equation in terms of a redefined field.
From the solution as a quantum state, we can extract classical information about the multi-point functions of the original velocity field.\footnote{
    Reference \cite{2026arXiv260110166G}, which shares similar spirit with us, has appeared during the revision process of this paper.}
Although there is a limitation in the information extraction step for turbulence with a high Reynolds number, our algorithm offers exponential advantages over classical methods in terms of the number of spatial grids $N_x$ (from ${\mathcal O}(N_x^2)$ to $\tilde{\mathcal O}(\mathrm{polylog} N_x)$).
This work serves as a proof of concept for solving nonlinear systems using quantum computing.

We leave it as a future research to ask whether going beyond the approximation \eqref{eq:approx_1st} to address a moderate or large Reynolds number is possible or not. We emphasize that, in our algorithm and its possible extensions, the question is how to deal with nonlinearity in the information extraction step.

A possible extension may be the inclusion of random forcing.
Burgers turbulence is not chaotic, unlike the full Navier--Stokes turbulence, and random initial conditions and/or random forcing have often been introduced in the study of the Burgers turbulence \cite{2000nlin.....12033F}.
It would be interesting to examine how a stochastic forcing term can be implemented in our approach.\footnote{
    This would prompt us to consider the Fokker--Planck approach, since it implements a Gaussian stochastic forcing term in general nonlinear equations of motion \cite{2024arXiv240113500T,2025RSPSA.48150016T}. We thank Luca Magri for bringing it to our attention.}

An even more ambitious future direction is to address a broader class of nonlinearities. Our method of applying a nonlinear transformation of variables may complement the variety of proposed linearization methods in the literature.
\\

\section*{acknowledgement}
\noindent
The work of FU is supported by JSPS KAKENHI Grant No.~JP23KJ0642 and by IBS under the project code IBS-R018-D3.
The work of KM is supported by MEXT Quantum Leap
Flagship Program (MEXT Q-LEAP) Grant no.
JPMXS0120319794, JSPS KAKENHI Grant No.
JP22K11924, and JST COI-NEXT Program Grant No.
JPMJPF2014.
The work of KF is supported by JSPS KAKENHI Grant No.~JP20K14512.\\

\appendix
\section{Perturbative flatness for a plane wave initial condition \label{appx:analytic}}
\noindent
For the plane-wave initial condition,
\begin{align}
    \psi(x,0)
        =1+\delta_m\sin\left(\dfrac{2\pi mx}{L}\right),\quad m\in{\mathbb Z}_+,
        \label{eq:plain wave}
\end{align}
one can analytically integrate Eq.~\eqref{eq:heateqn} to obtain
\begin{align}
    \psi(x,t)
        =1+\delta_m\sin\left(\dfrac{2\pi mx}{L}\right)\exp{\left(-\frac{4\pi^2m^2\tau}{N_x^2}\right)}.
\end{align}
In the large $\tau$ limit, it is clear that $\vert\vert\boldsymbol{\psi}_0\vert\vert/\vert\vert\boldsymbol{\psi}(\tau)\vert\vert=1$.

We can also easily compute $\beta$.
The velocity field, $u$, is computed by the relation \eqref{eq:psi_to_u}, and by expanding $\langle u^2\rangle$ and $\langle u^4\rangle$ by powers of $\delta_m$, one obtains
\begin{align}
    \langle u^2\rangle
    &=\left(\dfrac{4\pi m\nu\delta_m}{L}\right)^2\,\dfrac{1}{2}\,\exp\left(-\frac{8\pi^2m^2\tau}{N_x^2}\right)\notag\\
    &\quad\cdot\left(1+\dfrac{3\delta_m^2}{4}\,\exp\left(-\frac{8\pi^2m^2\tau}{N_x^2}\right)+\cdots\right),\\
    \langle u^4\rangle
    &=\left(\dfrac{4\pi m\nu\delta_m}{L}\right)^4\,\dfrac{3}{8}\,\exp\left(-\frac{16\pi^2m^2\tau}{N_x^2}\right)\notag\\
    &\quad\cdot\left(1+\dfrac{5\delta_m^2}{3}\,\exp\left(-\frac{8\pi^2m^2\tau}{N_x^2}\right)+\cdots\right).
\end{align}
By substituting it into Eq.~\eqref{eq:flatnessdef}, one obtains
\begin{align}
    \beta
        =-\dfrac{3}{2}\left(1-\dfrac{\delta_m^2}{6}\,\exp\left(-\frac{8\pi^2m^2\tau}{N_x^2}\right)+\cdots\right).
\end{align}

\section{Validity of the integration scheme}
\label{sec:demo}
\noindent It is important to test if the combination of transformation of variable and discretization of the spatial coordinate does not compromise the numerical solution so that it captures the characteristics of Burgers flows.
Here, we examine a shock-like solution and a rarefaction wave solution obtained by numerically integrating Eq.~\eqref{eq:discretized}.

\begin{figure}[ht]
        \begin{minipage}[h]{1.0\hsize}
            \includegraphics[keepaspectratio, width=0.96\textwidth]{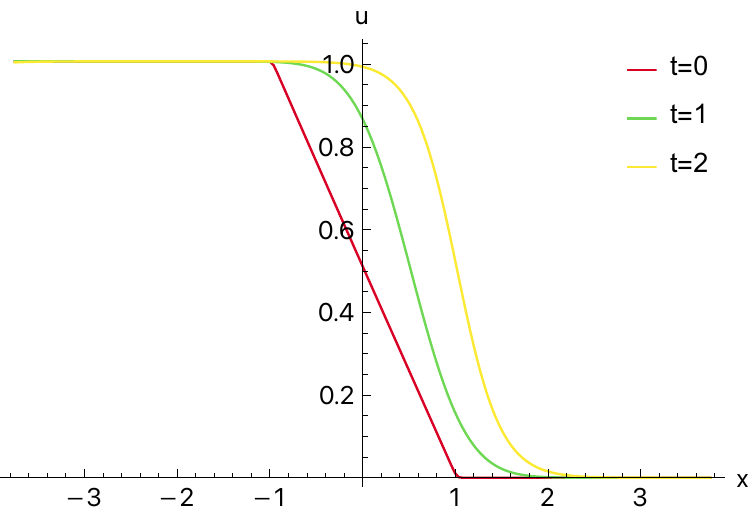}
        \end{minipage}
        \begin{minipage}[h]{1.0\hsize}
            \includegraphics[keepaspectratio, width=0.96\textwidth]{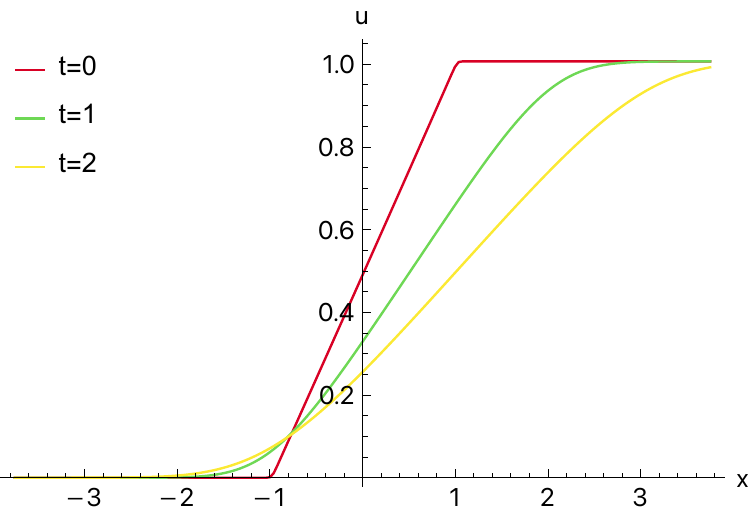}
        \end{minipage}
    \vspace{1mm}
    \caption{\label{fig:demo}
    Shock-like structure (Top) and rarefaction wave (Bottom).
    In each panel, solid lines correspond to solutions at $t=0$ (red), $t=1$ (green), and $t=2$ (yellow). 
    Our numerical code used to make this plot is public at \cite{mycode}.}
\end{figure}

For the calculations shown in Fig.~\ref{fig:demo}, we chose initial conditions such that a finite interval $\vert x\vert<1$ linearly interpolates $u=0$ and $u=1$, and $u$ stays constant outsider the interval.

When $u(x\leq-1)=1$ and $u(x\geq1)=0$ at the beginning (red line in the top panel), the solution develops a shock-like structure (green and yellow lines in the top panel).
Because of the finite viscosity, which we take to be $\nu=0.1$, the emerging structure is not a discontinuous shock but exhibits a smooth transition with a width $\sim \nu/u\sim\mathcal O(0.1)$.
The front of the transition moves rightward with velocity $(u(x\leq-1)+u(x\geq1))/2=0.5$.

When $u(x\leq-1)=0$ and $u(x\geq1)=1$ at the beginning (red line in the bottom panel), the velocity gradient gets shallower over time (green and yellow lines in the bottom panel).
The location of the center of the transition zone moves rightward with velocity $(u(x\leq-1)+u(x\geq1))/2=0.5$.

The test calculations show that our integration scheme applies well to the Burgers flows, apart from the approximation \eqref{eq:dpsi_to_u}, which is valid only when we have many of such structures as small-scale fluctuations on top of a background component $\tilde\psi$.
Note that,we did not impose periodic boundary conditions in the above test calculations, in order to highlight the behavior of the single structure around the origin.
We instead imposed the Neumann boundary condition, namely $\partial_x\psi=0$ at the boundary.

\bibliography{bib}
\end{document}